# Recovery of radioisotopes from nuclear waste for radio-scintillator-luminescence energy applications

**Alfred Bennun**[1,2]

[1]Graduate Programs in the Molecular Biosciences, Rutgers University, Newark, USA
[2]CONICET, Buenos Aires, Argentina
Email: alfr9@hotmail.com

## Abstract

Extraction of the light weight radioisotopes (LWR) $^{89}Sr/^{90}Sr$, from the expended nuclear bars in the Fukushima reactor, should have decreased the extent of contamination during the course of the accident. $^{89}Sr$ applications could pay for the extraction of $^{89}Sr/^{90}Sr$ from nuclear residues. Added value could be obtained by using $^{89}Sr$ for cancer treatments. Known technologies could be used to relate into innovative ways LWR, to obtain nuclear energy at battery scale. LWR interact by contact with scintillators converting β-radiation into light-energy. This would lead to manufacturing scintillator lamps which operate independently of other source of energy. These lamps could be used to generate photoelectric energy. Engineering of radioisotopes scintillator photovoltaic cells, would lead to devices without moving parts.

Key words: nano-scale battery, photoelectric energy, scintillator lamps, nuclear waste.

## Introduction

Fission plants use the reaction: $^{235}U^{92}+n \rightarrow {}^{90}Sr^3 + {}^{89}Sr^{38} + {}^{143}Xe^{54} + 3n + \gamma$, to generate electricity. $^{143}Xe$ ($t_{1/2}=30s$) $\rightarrow {}^{143}Cs + \beta \rightarrow {}^{142}Cs + n$. $^{89}Sr$ is a produced as a fission products in slightly less quantities than $^{90}Sr$, but their ratio is 166/1. $^{89}Sr$ has a specific activity of 28,200 Ci/g greater than 1g of radium.

In Fukushima, the nuclear plant left untreated the expended reactor bars [1] [2]. Hence, the heat generated by radioactivity, inside the expended bars was mainly that of $^{89}Sr$. The flooding from the reactor and by the sea dissolved these radioisotopes from the storage bars and carried then into the surrounding areas.

At 5 days after the accident, the measurement in Becquerel units (1 disintegration/s) the pollution in a neighboring town was found: 13-260Bq $^{89}Sr$ and $^{90}Sr$ 3.3-32Bq per kg of soil, amount strongly contaminating the cultivation areas. Radioactive Sr is extremely dangerous, because when ingested replaces $Ca^{2+}$ in their physiological functions. $^{90}Sr^{38}$ decay ($t_{1/2} = 28.8$ years) $\rightarrow {}^{90}Y^{39}$ ($t_{1/2}= 64hs$) + β (0.68MeV) + $\bar{v}$ $\rightarrow {}^{90}Zr^{40} + \beta$ (2.28MeV) + $\bar{v}$. While: $^{89}Sr^{38}$ ($t_{1/2}=50.55$ days) $\rightarrow {}^{90}Y^{39} + \beta(1.5MeV) + \bar{v}$.

A large emission number of low penetrating power particles in a short time characterize $^{89}Sr$, which allows that the highly radioactive could involve a rather limited danger.

The difference in $t_{1/2}$ allows $^{89}Sr$ to deliver its energy at a rate 200 times higher than $^{90}Sr$ and by decaying 99% in one year allows a left-over of uncontaminated $^{90}Sr$. This one becomes available for applications requiring long periods of time. β-radiation from $^{89}Sr$ damages animal tissues because ionize water, but penetrates through the skin about: 5 to 8 mm [3]. Thus, allowing manipulation of small amounts of $^{89}$Sr with gloves, glasses and thick laboratory clothing. In solution $^{89}$Sr has been administered as a palliative of pain of metastatic prostate cancer in doses of 150MBq [4]. Its bone-seeking properties favored its use at higher doses, in the treatment of bone cancer metastasis [5] [6] [7] [8] [9].

Evaluation of in-situ radiation treatments for replacing the therapeutic-radium on the treatment for prostate shrinking, skin cancer, etc., has involved many radioisotopes.

## Applications evaluation

In an insoluble form like $^{89}$Sr silicate, a solid, could be used in the form of near microscopic implants, that inserted in quantities related to the tumors size could have a slow but persistent effects on reducing the extend of metastasis.

β-particle energy dissipates through the 5eV stopping power per molecule of water. Since, 460KJ/mol is the average energy of ionization, the mean energy of 1MeV per each β−particle emitted by $^{89}$Sr, allows a theoretical maximal ionization: $1.8 \times 10^5$ molecules of water. Hence, 1nmol of pure $^{89}$Sr (89 ng equivalent to 95MBq) emits over 50 days $4 \times 10^{14}$ MeV β-particles, which multiply by $1.8 \times 10^5 = 7 \times 10^{19}$ ionized water molecules or $1.2 \times 10^{-1}$ mg of $H_2O$ or ml. This number corresponds to the size of a small tumor, but the radiation characteristics of the radioisotope allow the use of a large number of $^{89}$Sr-nano-implants for in-situ metastasis treatments.

$^{89}$Sr has a clear advantage over radium due to its high efficiency versus costs ratio, much less dangerous and by its short-lived decay could not involve high disposal expenses. However, the medical market is rather small for the overall quantities generated at nuclear facilities. However, if used as an alternative form of energy its consumption may match production and market offerings.

## Feasibility

Efficiency of photovoltaic cell (PV) has been highly improved and basically a LWR-lamp or radioisotope light generator (RLG) coupled to a PV system could be used to generate electricity: radioisotope-light-electric-generator (RLEG), and recharge batteries. The electricity generated in-situ allows a battery operated electric motor which could not require fuel and the engineering of optical fibers permits transmission of light power.

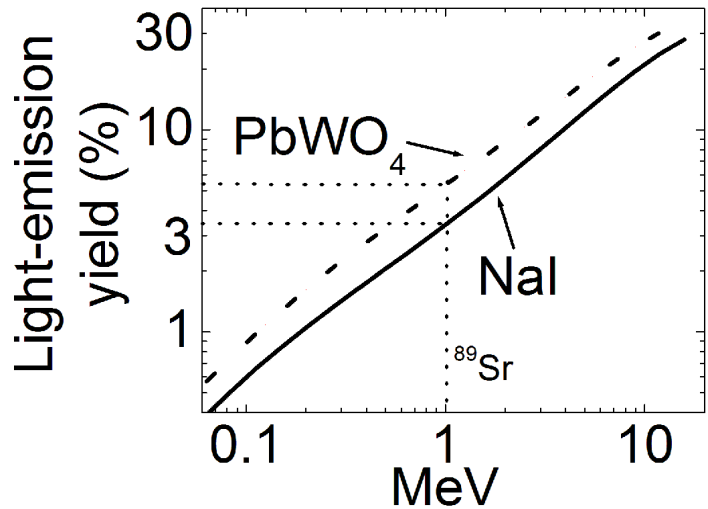

**Fig. 1. Performance-scintillator (η) vs. -Particle energy (MeV).** Radioactive light generators (RLG) are based in the scintillator stopping power capable of exciting an orbital with photon emission. **NaI** (3.675 g/cm$^3$), average energy of excitation (a.e.e.) 458.0 eV, 2% Tl + β [1MeV] → h$\nu$ (u.v.), η=3.4%; **PbWO$_4$** (Mw=455g/mol at 8.28 g/cm$^3$), a.e.e.: 616.7 eV, 2% Tl+ β [1MeV] → h$\nu$ (u.v.), η=5.4%.

The scintillator-crystals act as a shield by absorbing the kinetic energy of β-particles to reach an excited quantum state and return to its initial state through photon emission. For large quantities, it could be used unbreakable glass as an additional covering shield because it is transparent to light.

The silicates are materials based on the repeating unit SiO$_4^{4-}$ tetrahedral. SiO$_4^{4-}$ unit contains negative charges which are generally compensated by the presence of alkali metal ions like $^{89}$Sr recovered and purified from mixed uranium fission products by dissolving with HCl. The $^{89}$SrCl$_2$ titrated to the required pH to form $^{89}$Sr(HO)$_2$ could then by processed by mixing with the dissolution of potassium/sodium silicate K/NaSiO$_3$ to obtain tetrahedral crystals of $^{89}$SrSiO$_4$.

The silicate dissolution in water in the proportion of the molecular relationship between n=0.5 and n=4: nSiO$_2$+Na$_2$O (K$_2$O) are viscose liquids, with an alkaline increase of pH from 10 to 14, as a function of decreasing the value of n.

Wolfram W$^{74}$ also known as tungsten, with a melting point of 3,410°C, has a high traction resistance, very low vapor pressure and from all the metals is the one with lower dilatation coeficient. Since W$^{74}$ participates in the molecular structure of as several scintillators like PbWO$_4$, the metal provide the advantage to form higly temperature and mechanical resistant crystals.

Tungstate in alkaline solutions depolymerizes to WO$_4^{2-}$. Increasing molar ratio of [WO$_4^{2-}$] to [$^{89}$Sr$^{2+}$] modifies the morphology of the synthesized crystals of $^{89}$SrWO$_4$ [11] [12] [13] from rods to other structures like dumbbells, or notched spheres. It could predicted that Tl$^+$-doped RLG-chip design of $^{89}$SrWO$_4$ , could have as a scintillator a 3.8-4.4%, light emission yield.

Crystals of $^{89}$SrWO$_4$ could be used instead of electricity, as the energy source to elicit the phosphor response of PbWO$_4$, which if incident in electrovoltaic cells may leads to an additional photocurrent response of PbWO$_4$. The latter, has a green luminescent response at low temperature, and has been used to manufacture electrically swtched light bulbes [12] [13]. Hence, PbWO$_4$ could be used as a covering to increasing light emission yield of $^{89}$SrWO$_4$, which as a 87mg-Chip, this will equiparate with a continuous electrical output consumption of 15Watt for over 50days. Higher manufacturing costs would be compensate by nill consuption costs.

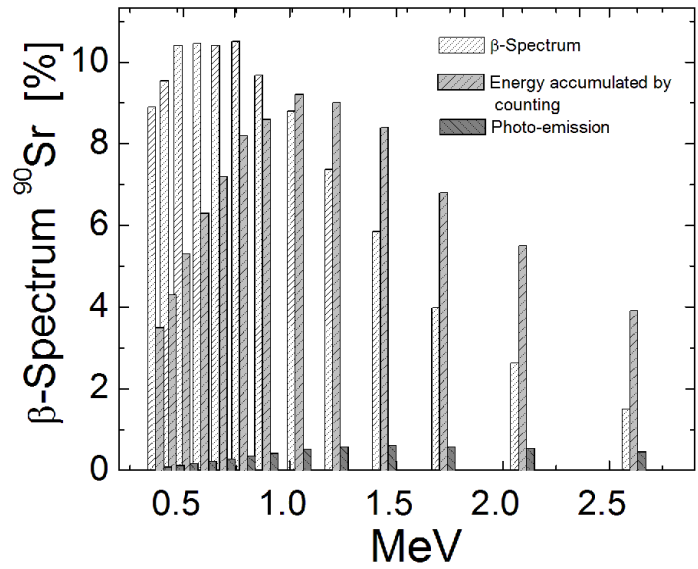

**Fig. 2. β-Energy distribution.** Calculated as a percentege of the value obtained by suming-up the energy of β-particles in the range between 0.39MeV and 2.6MeV (▨). The percentage of β-particles with less than 0.6 MeV constitute 40% of the total, but sum up to less than 20% of total energy. The particles emitted at higher than 0.6 MeV, constitute 60% with 80% of total energy emitted at an average 1.4MeV (▨). The scintillator range efficiency for aproach 7% for PbWO$_4$ (▨).

The fig.2 indicates the convenience of using a scintillator mixture to obtain a better response for the β emitted at lower energy and ion dotting of the crystal like that of BaF$_2$:PbWO$_4$ to increase maximal transmittance [14].

The scintillator POPOP which can be used solid or liquid as secondary photon emission (λ), also extends λ-emission to 410nm in the violet of the visible spectrum. Crystals with added Al$_2$O$_3$ optimize light transmission and give a red shift of the emission spectrum. The omission of water in the engineering of the RLG crystals may decrease efficiency losses. $^{89}$Sr light emitting panels allow an electric energy independent source of cheap energy to operate greenhouses located in Polar Regions. These radioisotope photo-cells emit light 24h without interruption, an engineering solution could be that of a rotating system which by displacement over 120 degrees, can illuminate every 8hs a different area, triplicating harvest yields. Alternatively, the $^{89}$Sr light emitting could be re-directed some part of the day to activate photovoltaic cells.

$^{89}$Sr in amounts smaller than 1g do not require active cooling [15][16][17][18], larger amounts could be ventilated, and the radio-luminescent cell could be manufactured as a fiber around a coolant tube, surrounded by mirrors pointing toward the photovoltaic

cells. Engineering of more than one scintillator in layers plus their geometrical arrangements around the $^{89}$Sr-core adding to an adequate arrangement of mirrors could be investigated to improve system efficiency. $^{89}$Sr core of a battery can be contained in a metal box with external plugs. Only qualified personnel may be provided with an access key. Also may be convenient that from the site for radioisotope generated light, could be, by using optical fibers, transmitted-out to activate elsewhere light consuming devices.

*Dr. Alfred Bennun.*
*Full Professor- Rutgers University (R)*
*Graduate Programs in the Molecular BioSciences- Rutgers-UMDNJ*
*Elected member Nuclear American Society – Radioisotopes and Education Division*
*Graduated Nuclear Expert by the Advanced Study Program for the use of radioisotopes in Biology, dictated by J.A.B. Cooper and M. Lipton, Northwestern University, U.B.A. and the U.S. department of State,*
*http://www.ncbi.nlm.nih.gov/pubmed?term=Bennun%20A.*
*http://www.biomedexperts.com*
*Godoy Cruz 3046, Torre 2, 8° Polo. Ciudad de Buenos Aires, CP: 1425*
*alfr9@hotmail.com*